\documentclass[epj,final]{svjour}
\usepackage{graphicx}
\begin{document}
\title{Relation between Vortex Excitation and Thermal Conductivity 
in Superconductors}
\author{Mitsuaki Takigawa\thanks{\email{takigawa@mp.okayama-u.ac.jp}}, %
Masanori Ichioka and Kazushige Machida  
}                     
\institute{Department of Physics,%
        Okayama University, Okayama 700-8530, Japan}
\abstract{
Thermal conductivity $\kappa_{xx}(T)$ under a field is investigated 
in $d_{x^2-y^2}$-wave superconductors and isotropic $s$-wave 
superconductors  by the linear response theory, using a microscopic 
wave function of the vortex lattice states. 
To study the origin of the different field dependence of 
$\kappa_{xx}(T)$ between higher and lower temperature regions, 
we analyze the spatially-resolved thermal conductivity around a vortex 
at each temperature, which is related to the spectrum of the local density 
of states. 
We also discuss the electric conductivity in the same formulation 
for a comparison. 
} 
\maketitle
%

\section{Introduction}
\label{intro}

Recent advance to synthesize new exotic superconducting
materials further requires the experimental probes to
precisely identify their pairing functions consisting of
the orbital and spin components. The orbital part of the
pairing function determines the nodal structure of the energy gap
on the Fermi surface.
Thermal conductivity is one of the standard techniques
to probe the node of the gap structure
~\cite{Lussier,Izawa,Nishira,Hirschfeld1996,Schmitt,Suderow}. 
One can basically extract the gap topology such as line or point nodes
by analyzing its dependence on the temperature $T$.

In the vortex state under a magnetic field, 
the thermal conductivity is affected by the low energy quasiparticle 
state around the vortex~\cite{Suderow,Sousa,Aubin1999}.
So far, the thermal conductivity in the vortex state has been investigated 
by the theory of gapless superconductors at high field~\cite{Houghton,Yin}, 
or the theory considering vortices as scattering centers~\cite{Cleary}.
Recently, the thermal conductivity in the vortex state of high-$T_{\rm c}$ 
superconductors attracts much attention, because the quasiparticle states 
are qualitatively different in the vortex state between the $d$-wave 
pairing case and the conventional $s$-wave pairing cases. 
In the $s$-wave pairing, low energy quasiparticle states are bounded around 
the vortex core~\cite{CdGM,HayashiPRL}. 
In the $d$-wave pairing, low energy quasiparticle states around the 
vortex extend outside of the vortex core due to the line node of the 
superconducting gap%
~\cite{Volovik,Ichioka96,Ichioka99A,Ichioka99B,FranzBdG,Franz2000}. 
The contribution to the thermal conductivity from the quasiparticles 
outside of the core is investigated by the theory of the Doppler shift 
(or the Dirac fermion) in the $d$-wave pairing 
case~\cite{Kubert,Vekhter,Franz01}. 
Since these theory neglect the contribution from the quasiparticle 
within the vortex core~\cite{Volovik}, 
they are not applied to the $s$-wave superconductors.

Here, we calculate the $T$-dependence of the thermal conductivity 
under a field. 
Our calculation is based on a microscopic theory 
of Bogoliubov-de Gennes (BdG) equation for describing the vortex 
state~\cite{Wang,TakigawaPRL,TakigawaJPSJ} and 
standard linear response theory for the thermal 
conductivity~\cite{Kadanoff,Ambegaokar}, 
assuming a clean-limit type II superconductors.
Our calculation includes all contributions from the inside and the outside  
region of the vortex core. 
The spatial distribution of the thermal conductivity is calculated 
from the wave function of the vortex lattice state, and analyzed by 
compared with the spatial distribution of the local density of 
states (LDOS). 

The rest of this paper is organized as follows. 
In Sec. \ref{Form}, we describe our formulation based on the BdG equation 
and the linear response theory. 
In Sec. \ref{Thermal}, we study the dependence of the 
thermal conductivity on the temperature in the vortex state. 
We also show the position-resolved thermal conductivity and 
its energy decomposition, 
and discuss the relation with the LDOS. 
In Sec. \ref{El-Cond}, we investigate the electric conductivity  
in the same formulation, and discuss the difference of the quasiparticle 
contribution between the thermal conductivity and electric conductivity. 
The summary and discussions are given in Sec. \ref{summary}. 

\section{Formulation}
\label{Form}

\subsection{Bogoliubov-de Gennes equation}
\label{sec:subBdG}

We obtain the wave function in the vortex lattice state by solving 
the BdG equation 
for the extended Hubbard model~\cite{Wang,TakigawaPRL,TakigawaJPSJ}
defined on a two dimensional square lattice.
From this model, we obtain qualitatively the same quasi particle 
structure as previous theoretical studies 
both for the $s$-wave~\cite{CdGM,HayashiPRL} case 
and for the $d$-wave~\cite{Volovik,Ichioka96,Ichioka99A,Ichioka99B,FranzBdG,Franz2000} case.
Here we briefly discuss the BdG equation for the $s$-wave pairing and 
the $d$-wave pairing cases, which is written as
\begin{equation}
\sum_j
\left( \begin{array}{cc}
K_{i,j} & D_{i,j} \\ D^\dagger_{i,j} & -K^\ast_{i,j}
\end{array} \right)
\left( \begin{array}{c} u_\alpha({\bf r}_j) \\ v_\alpha({\bf r}_j)
\end{array}\right)
=E_\alpha
\left( \begin{array}{c} u_\alpha({\bf r}_i) \\ v_\alpha({\bf r}_i)
\end{array}\right) ,
\label{eq:BdG1}
\end{equation}
where
\begin{eqnarray} &&
K_{i,j}=-\tilde{t}_{i,j}  - \delta_{i,j}\mu, 
\label{eq:BdG2}\\ &&
\tilde{t}_{i,j}=t_{i,j} \exp [ {\rm i}\frac{\pi}{\phi_0}\int_{{\bf
r}_i}^{{\bf r}_j} {\bf A}({\bf r}) \cdot d{\bf r} ] , 
\label{eq:BdG2b}\\ &&
D_{i,j}=\delta_{i,j} U \Delta_{i,i} + \frac{1}{2}V_{i,j}\Delta_{i,j}
\label{eq:BdG3}
\end{eqnarray}
with the on-site interaction $U$, the flux quantum $\phi_0$ and the chemical
potential $\mu$.
The nearest neighbor (NN) transfer integral $t_{i,j}=t$
and the NN interaction $V_{i,j}=V$
for the NN site pair ${\bf r}_i$ and ${\bf r}_{i \pm {\hat e}}$.
The vector potential ${\bf A}({\bf r})=\frac{1}{2}{\bf H}\times{\bf r}$ 
in the symmetric gauge.
Since we assume an extreme type-II superconductor, 
the internal field term of ${\bf A}({\bf r})$ is neglected. 
The self-consistent condition for the pair potential is
\begin{equation}
\Delta_{i,j}=-\frac{1}{2}\sum_\alpha u_\alpha({\bf r}_i)
v^\ast_\alpha({\bf r}_j) \tanh(E_\alpha /2T) .
\label{eq:BdGsc}
\end{equation}
The $s$-wave pair potential is given by
\begin{equation}
\Delta_s({\bf r}_i)=U\Delta_{i,i}.
\label{eq:sOP1}
\end{equation}
The $d_{x^2-y^2}$-wave pair potential is
\begin{equation}
\Delta_{d}({\bf r}_i)=\frac{V}{4}(\Delta_{\hat{x},i} + \Delta_{-\hat{x},i}
- \Delta_{\hat{y},i} - \Delta_{-\hat{y},i} )
\label{eq:dOP1}
\end{equation}
with
\begin{equation}
\Delta_{\pm\hat{e},i}=\Delta_{i,i \pm \hat{e}}
\exp[{\rm i}\frac{\pi}{\phi_0}
\int_{{\bf r}_i}^{({\bf r}_i+{\bf r}_{i \pm \hat{e}})/2}
{\bf A}({\bf r}) \cdot d{\bf r}].
\label{eq:dOP2}
\end{equation}

We study the case of the square vortex lattice where the NN vortex is
located in the direction of $45^\circ$ from the $a$-axis.
This vortex lattice configuration is suggested for $d$-wave
superconductors and  $s$-wave superconductors with fourfold symmetric
Fermi surface~\cite{Ichioka99B,IchiokaGL,DeWilde,Kogan,Won}. 
The unit cell in our calculation is the square area of
{$N_r$}$^2$ sites where two vortices are accommodated.
Then, $H = 2\phi_0 / (a{N_r})^2$ with the lattice constant $a$.  
Thus, we denote the  field strength by $N_r$ as $H_{N_{r}}$.
Since $H$ should be commensurate with the atomic lattice, 
our formulation does not treat the field dependence continuously. 
We consider the area of {$N_k$}$^2$ unit cells.
By introducing the quasi-momentum of the magnetic Bloch state,
${\bf k}=(2\pi /aN_rN_k)(l_x,l_y):(l_x,l_y = 1,2,\cdots,N_k)$,
we set
$u_\alpha({\bf r})=\tilde{u}_\alpha({\bf r})
{\rm e}^{{\rm i} {\bf k}\cdot{\bf r}},
v_\alpha({\bf r})=\tilde{v}_\alpha({\bf r})
{\rm e}^{{\rm i} {\bf k}\cdot{\bf r}}$.
Then, the eigen-state of $\alpha$ is labeled by {\bf k} and 
the eigenvalues obtained by this calculation within a unit cell.

The periodic boundary condition is given by the symmetry
for the translation ${\bf R}=l_x{\bf R}_x^0 + l_y {\bf R}_y^0$, where 
${\bf R}_x^0=(a N_r,0)$ and ${\bf R}_y^0=(0,a N_r)$ are unit vectors 
of the unit cell for our calculation. 
Then, the translational relation is given by 
$\tilde{u}_\alpha({\bf r}+{\bf
R})=\tilde{u}_\alpha({\bf r}) {\rm e}^{i\chi({\bf r},{\bf R})/2},
\tilde{v}_\alpha({\bf r}+{\bf R})=\tilde{v}_\alpha({\bf r})
{\rm e}^{-i\chi({\bf r},{\bf R})/2}$.
Here,
\begin{equation}
\chi({\bf r},{\bf R})
= -\frac{2\pi}{\phi_0}{\bf A}({\bf R})\cdot{\bf r}
- 2\pi l_x(l_x-l_y) + \frac{2 \pi}{\phi_0}
({\bf H}\times {\bf r}_0)\cdot{\bf R}
\end{equation}
in the symmetric gauge. 
The vortex center is located
at ${\bf r}_0+\frac{1}{4}(3{\bf R}_x^0+{\bf R}_y^0)$.
The phase factor in Eq. (\ref{eq:dOP2}) is necessary 
to satisfy the translational relation
$\Delta_{d}({\bf r}+{\bf R})
=\Delta_{d}({\bf r}){\rm e}^{{\rm i}\chi({\bf r},{\bf R})}$.

The following parameter values are chosen in the calculation. 
The average electron density per site $\sim 0.9$ by
appropriately adjusting the chemical potential $\mu$.
We normalize all the energy scales by the
transfer integral $t$.
For the $s$-wave case, we set $U=-2.32t$ and $V=0$.
The resulting order parameter $\Delta_0=0.5t$ at $T=H=0$, 
and the superconducting transition temperature $T_c$ $\sim$ $0.27t$. 
For the $d$-wave case, we set $U=0$ and $V=-4.2t$. 
Then, $\Delta_0=1.0t$, and $T_c$ $\sim$ $0.42t$.
Our results do not qualitatively depend on the choice of these parameters.

\subsection{Local density of states}
\label{sec:LDOS}
In order to calculate physical quantities, we must construct 
the Green's functions from 
$E_{\alpha}$,$u_{\alpha}({\bf r})$,$v_{\alpha}({\bf r})$ 
in the formulation of imaginary time $\tau$ 
and Fermionic Matsubara frequency $\omega_n=2\pi T(n+\frac{1}{2})$.
The Matsubara Green's functions is given by 
\begin{equation} 
\hat{g}({\bf r},{\bf r}',{\rm i}\omega_n)= 
\left(\begin{array}{cc}
g_{11}({\bf r},{\bf r}',{\rm i}\omega_n) & 
g_{12}({\bf r},{\bf r}',{\rm i}\omega_n) \\ 
g_{21}({\bf r},{\bf r}',{\rm i}\omega_n) & 
g_{22}({\bf r},{\bf r}',{\rm i}\omega_n)  \end{array} 
\right),
\label{eq:gm}
\end{equation} 
where matrix components are Fourier transformation of 
\begin{eqnarray} 
g_{11}({\bf r},\tau,{\bf r}',\tau') =
-\langle T_{\tau}[ \psi_{\uparrow}({\bf r},\tau) 
                   \psi^\dagger_{\uparrow}({\bf r}',\tau')]\rangle , \\ 
g_{12}({\bf r},\tau,{\bf r}',\tau') =
-\langle T_{\tau}[ \psi_{\uparrow}({\bf r},\tau) 
                   \psi_{\downarrow}({\bf r}',\tau')]\rangle , \\
g_{21}({\bf r},\tau,{\bf r}',\tau') =
-\langle T_{\tau}[ \psi^\dagger_{\downarrow}({\bf r},\tau) 
                   \psi^\dagger_{\uparrow}({\bf r}',\tau')]\rangle , \\ 
g_{22}({\bf r},\tau,{\bf r}',\tau') =
-\langle T_{\tau}[ \psi^\dagger_{\downarrow}({\bf r},\tau) 
                   \psi_{\downarrow}({\bf r}',\tau')]\rangle . 
\end{eqnarray}  
The Green's functions in Eq. (\ref{eq:gm}) as follows,\cite{TakigawaJPSJ}
\begin{eqnarray} 
g_{11}({\bf r},{\bf r}',{\rm i}\omega_n) =&&
\sum_{\alpha}\frac{u_{\alpha}({\bf r})u_{\alpha}^{*}({\bf r}')}
{{\rm i}\omega_n - E_{\alpha}},
\label{eq:gm1} \\
g_{12}({\bf r},{\bf r}',{\rm i}\omega_n) =&&
\sum_{\alpha}\frac{u_{\alpha}({\bf r})v_{\alpha}^{*}({\bf r}')}
{{\rm i}\omega_n - E_{\alpha}},
\label{eq:gm2} \\
g_{21}({\bf r},{\bf r}',{\rm i}\omega_n) =&&
\sum_{\alpha}\frac{v_{\alpha}({\bf r})u_{\alpha}^{*}({\bf r}')}
{{\rm i}\omega_n - E_{\alpha}},
\label{eq:gm3} \\
g_{22}({\bf r},{\bf r}',{\rm i}\omega_n) =&&
\sum_{\alpha}\frac{v_{\alpha}({\bf r})v_{\alpha}^{*}({\bf r}')}
{{\rm i}\omega_n - E_{\alpha}}. 
\label{eq:gm4}
\end{eqnarray}

From the LDOS is given by the thermal Green's functions as 
\begin{eqnarray}
N_\uparrow(E,{\bf r})&&
=-{1 \over \pi}{\rm Im} g_{11}({\bf r},{\bf r},
{\rm i}\omega_n \rightarrow E+{\rm i}0^+)
\nonumber \\
&&=\sum_\alpha |u_\alpha({\bf r})|^2 \delta(E-E_\alpha)
\end{eqnarray}
for the up-spin electron contributions, and 
\begin{eqnarray}
N_\downarrow(E,{\bf r})&&
={1 \over \pi}{\rm Im} g_{22}({\bf r},{\bf r},
-{\rm i}\omega_n \rightarrow E+{\rm i}0^+)
\nonumber \\
&&=\sum_\alpha |v_\alpha({\bf r})|^2 \delta(E+E_\alpha)
\end{eqnarray}
for the down-spin electron contributions.
Therefor, the LDOS is given by 
\begin{eqnarray}
N(E,{\bf r})&&=N_\uparrow(E,{\bf r})+N_\downarrow(E,{\bf r}) 
\nonumber \\
&&=\sum_\alpha \{ |u_\alpha({\bf r})|^2 \delta(E-E_\alpha) 
+ |v_\alpha({\bf r})|^2 \delta(E+E_\alpha) \}.
\nonumber \\
\end{eqnarray}

\subsection{Linear response theory}
\label{Linear}

We calculate the thermal conductivity following the method of Refs. 
\cite{Kadanoff} and \cite{Ambegaokar}.  
According to the linear response theory,
thermal current $h_x({\bf r}_1)$ flowing to the $x$-direction at 
${\bf r}_1$-site is given by 
\begin{eqnarray}
&&h_x({\bf r}_1) =
\nonumber \\
&&\frac{1}{T}\sum_{{\bf r}_2} {\rm Re}\{%
{1\over {\rm i}}\frac{\rm d}{{\rm d} \Omega} 
Q_{xx}({\bf r}_1,{\bf r}_2,{\rm i}\Omega_n\rightarrow\Omega+{\rm i} 0^+ ) 
\}_{\Omega\rightarrow 0} %
\nonumber \\
&&\times (-\nabla_xT({{\bf r}_2})),
\end{eqnarray}
when the small temperature gradient $-\nabla_xT({\bf r}_2)$ 
along the $x$-direction is applied at ${\bf r}_2$-site. 
The heat-heat correlation function is defined by 
\begin{eqnarray}&&%
Q_{xx}({\bf r}_1 \tau_1,{\bf r}_2 \tau_2)%
=\langle T_{\tau}[h_x({\bf r}_1 \tau_1),h_x({\bf r}_2 \tau_2)]\rangle
\nonumber \\
&&=T\sum_{n}{\rm e}^{-{\rm i}\Omega_n(\tau_1-\tau_2)}
Q_{xx}({\bf r}_1,{\bf r}_2,{\rm i}\Omega_n) 
\label{eq:Qxx}
\end{eqnarray}
in the formulation of imaginary time $\tau$ and 
Matsubara frequency  $\Omega_n$. 
The heat current operator ${\bf h}({{\bf r}}_j,\tau)$ is written as
\begin{eqnarray}
{\bf h}({\bf r}_j,\tau) &=& 
-\frac{\rm i}{2m} \left( {\bf P}_j \frac{\partial}{\partial\tau'} 
-{\bf P}_{j'}^{\dagger} \frac{\partial}{\partial\tau} \right)
\nonumber \\ && 
\times
\sum_{\sigma}\psi^{\dagger}_{\sigma}({\bf r}_{j'},\tau') 
\psi_{\sigma}({\bf r}_j,\tau)|_{j=j',\tau=\tau'}
\end{eqnarray}
in terms of the electron field operators $\psi_{\sigma}({\bf r}_j,\tau)$. 
The $x$-component of the momentum operator ${\bf P}_{j}$ in 
the discretized square lattice is defined as 
\begin{equation} 
[{\bf P}_{j}\psi_{\sigma}({\bf r}_j,\tau)]_x ={\bf a} \frac{2mt}{\rm i}
{\rm e}^{{\rm i}(\pi / \phi_0){\bf a}\cdot
{\bf A}({{\bf r}}_{j}+{\bf a}/2)}\psi_{\sigma}({\bf r}_{j}+{\bf a},\tau)
\end{equation} 
with ${\bf a}=(a,0)$. 
Four electron field operators in Eq. (\ref{eq:Qxx}) are 
decomposed by using the Matsubara Green's functions 
in Eqs. (\ref{eq:gm1})-(\ref{eq:gm4}) .

For the study of the thermal conductivity, we have to introduce the 
dissipation term $\eta$ in the Green's function. 
Then, the retarded and 
advanced Green's functions are, respectively, given by 
$\hat{G}^{\rm R}({\bf r},{\bf r}',\omega)= 
\hat{g}({\bf r},{\bf r}',{\rm i}\omega_n \rightarrow \omega+{\rm i}\eta)$ and 
$\hat{G}^{\rm A}({\bf r},{\bf r}',\omega)= 
\hat{g}({\bf r},{\bf r}',{\rm i}\omega_n \rightarrow \omega-{\rm i}\eta)$. 
Therefore, in the spectral representation, we obtain 
\begin{equation} 
\hat{g}({\bf r},{\bf r}',{\rm i}\omega_n)= \int_{-\infty}^\infty 
\frac{{\rm d}\omega}{2\pi}\frac{\hat{A}({\bf r},{\bf r}',\omega)}
{{\rm i}\omega_n -\omega} , 
\end{equation} 
where 
\begin{eqnarray} && 
\hat{A}({\bf r},{\bf r}',\omega) 
=-\frac{1}{2 \pi {\rm i}} \left( 
 \hat{G}^{\rm R}({\bf r},{\bf r}',\omega)
-\hat{G}^{\rm A}({\bf r},{\bf r}',\omega) \right) 
\nonumber \\ &&  
=\sum_\alpha \delta_\eta(\omega-E_\alpha) 
\left(\begin{array}{cc} 
u_\alpha({\bf r}) u^\ast_\alpha({\bf r}') & 
u_\alpha({\bf r}) v^\ast_\alpha({\bf r}') \\ 
v_\alpha({\bf r}) u^\ast_\alpha({\bf r}') & 
v_\alpha({\bf r}) v^\ast_\alpha({\bf r}')  \end{array} 
\right) 
\end{eqnarray} 
with $\delta_\eta(\omega)=\eta [\pi (\omega^2+\eta^2)]^{-1}$.

As a result, the heat-heat correlation function is reduced to
\begin{eqnarray} &&
\left. 
\frac{{\rm d}}{{\rm d}\Omega}Q_{{\bf P}_1{\bf P}_2}
({\bf r}_1,{\bf r}_2,{\rm i}\Omega_n \rightarrow \Omega+{\rm i}0^+) 
\right|_{\Omega=0}
\nonumber \\ &&
=\frac{1}{4m^2}\sum_{\alpha\beta}F({E_\alpha,E_\beta})
\nonumber \\ &&
\times
[{{\bf P}_1u_{\alpha}({\bf r}_1)u_{\beta}^\ast ({\bf r}_{1'})
+u_{\alpha}({\bf r}_1){\bf P}_{1'}^{\dagger}u_{\beta}^\ast ({\bf r}_{1'})}
\nonumber \\ &&
+{{\bf P}_1^{\dagger}v_{\alpha}({\bf r}_1)v_{\beta}^\ast ({\bf r}_{1'})
+v_{\alpha}({\bf r}_1){\bf P}_{1'}v_{\beta}^\ast ({\bf r}_{1'})}]
\nonumber \\ &&
\times
[{{\bf P}_2u_{\alpha}({\bf r}_2)u_{\beta}^\ast ({\bf r}_{2'})
+u_{\alpha}({\bf r}_2){\bf P}_{2'}^{\dagger}u_{\beta}^\ast ({\bf r}_{2'})}
\nonumber \\ &&
+{{\bf P}_2^{\dagger}v_{\alpha}({\bf r}_2)v_{\beta}^\ast ({\bf r}_{2'})
+v_{\alpha}({\bf r}_2){\bf P}_{2'}v_{\beta}^\ast ({\bf r}_{2'})}]|_{1=1',2=2'},
\label{eq:QPP}
\end{eqnarray}
where
\begin{eqnarray} &&
F({E_\alpha,E_\beta})
\nonumber \\ &&
=\int\frac{{\rm d}\omega}{2\pi}\int\frac{{\rm d}\omega'}{2\pi}
\delta_{\eta}(\omega-E_\alpha)\delta_{\eta}(\omega'-E_\beta)
\nonumber \\ &&
\times
[{\rm P}\frac{\omega^2f(\omega)-\omega'^2f(\omega')}{(\omega-\omega')^2}
+{\rm i}\pi\omega^2f'(\omega)\delta(\omega-\omega')].
\label{eq:Fab0}
\end{eqnarray}

When the temperature gradient $\nabla_xT({\bf r}_2)$ is uniform, 
the position-resolved thermal conductivity is written as
\begin{eqnarray} && 
\kappa_{xx}({\bf r}_1)
=\frac{h_x({\bf r}_1)}{-\nabla_xT} 
\nonumber \\ && 
=\frac{1}{T}{\rm Im}\{
\frac{\rm d}{{\rm d}\Omega} \frac{1}{N} \sum_{{\bf r}_2} Q_{xx}({\bf r}_1, 
{\bf r}_2, {\rm i}\Omega_n\rightarrow\Omega+{\rm i}0^+ )\}.  
\label{eq:kappar}
\end{eqnarray}
The spatially averaged thermal conductivity
\begin{equation}
\kappa_{xx}=\frac{1}{N}\sum_{{\bf r}_1}\kappa_{xx}({\bf r}_1)
\end{equation}
is observed in the experiment. $N$ is the total number of sites. 
At zero field, our formulation for $\kappa_{xx}$ is reduced to the 
well-known formula for the uniform superconductors 
in the presence of impurity scattering~\cite{Kadanoff}. 

Using the wave function $u_\alpha({\bf r})$, 
$v_\alpha({\bf r})$ and the eigen-energy $E_\alpha$ in Eq. (\ref{eq:BdG1}), 
we calculate the dependence of $\kappa_{xx}$ on the temperature and the field. 
And to analyze this behavior, we also study the spatial structure of 
$\kappa_{xx}({\bf r})$, i.e., the local contribution to $\kappa_{xx}$, 
Here, we neglect the principal value integral term 
${\rm Re}F(E_\alpha,E_\beta)$ in Eq. (\ref{eq:Fab0}),
because the contribution from this term vanishes in the spatial average 
of $\kappa_{xx}({\bf r})$. 
We typically choose $\eta = 0.01t$.
\begin{figure*}[t]
\center{\includegraphics[]{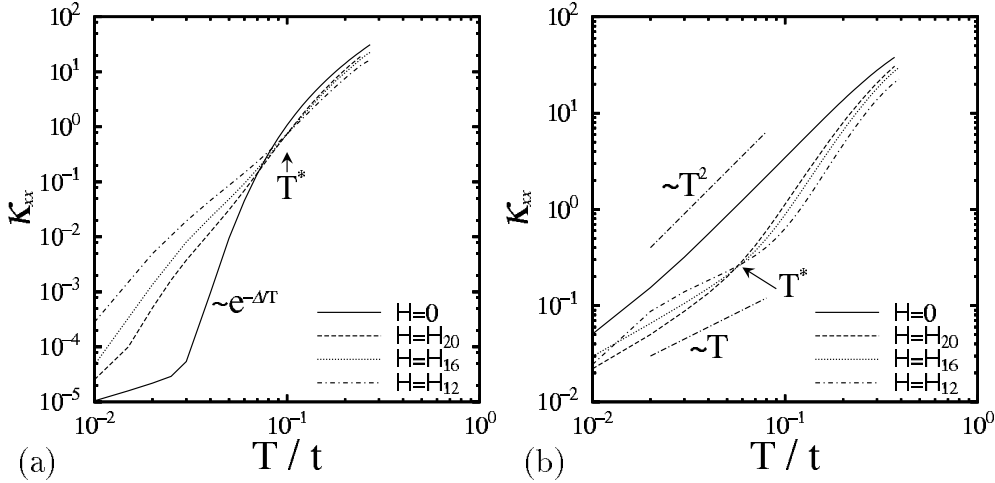}}
\caption{\label{fig:k-T}
Temperature dependence of $\kappa_{xx}(T)$ 
at $H=0$, $H_{20}$, $H_{16}$ and $H_{12}$. 
(a) $s$-wave case. (b) $d$-wave case.  
}
\end{figure*}
%

\section{Thermal Conductivity}
\label{Thermal} 
\subsection{Temperature dependence of thermal conductivity}

\begin{figure*}
\center{\includegraphics[]{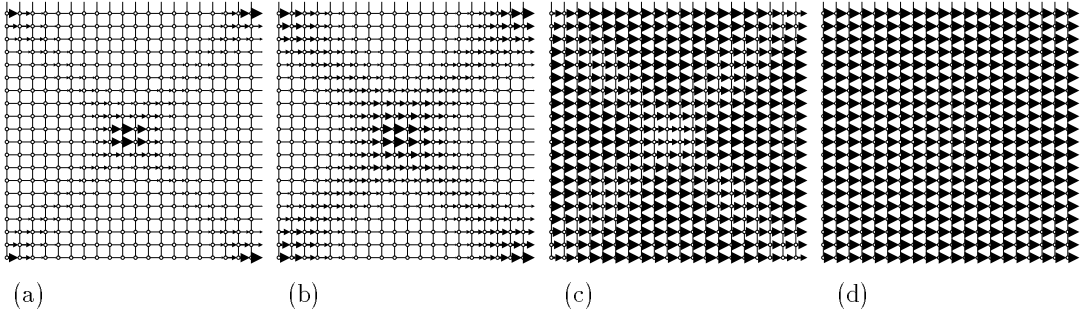}}
\caption{\label{fig:k-r-s}
The vector plots of the $\kappa_{xx}({\bf r})$ in the $s$-wave pairing case %
at $T/t$=0.02 (a), 0.05 (b), 0.09 (c) and 0.13 (d). %
$H=H_{20}$. 
The vortices are located at the center %
and the four corners in the figure. %
Arrow size is  proportional to flow strength.%
}
\end{figure*}

\begin{figure*}
\center{\includegraphics[]{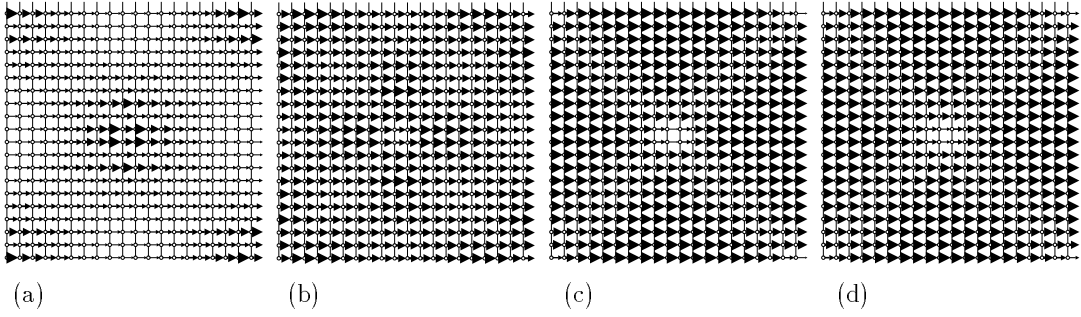}}
\caption{\label{fig:k-r-d}
The vector plots of the $\kappa_{xx}({\bf r})$ in the $d$-wave pairing case %
at $T/t$=0.02 (a), 0.05 (b), 0.09 (c) and 0.13 (d). %
$H=H_{20}$.  
The vortices are located at the center %
and the four corners in the figure. %
Arrow size is  proportional to flow strength.%
}
\end{figure*}

The $T$-dependence of $\kappa_{xx}(T)$ is shown 
in Figs. \ref{fig:k-T}(a) and (b) for the $s$-wave and the $d$-wave 
pairing cases, respectively. 
For the $s$-wave pairing, it shows exponential $T$-dependence 
due to the full gap of the $s$-wave superconductivity at $H=0$. 
It changes into a $T$-linear behavior at low $T$ region in the vortex 
state at $H\neq0$, reflecting low energy quasiparticle state 
around the vortex. 
We see also that the deviation from the expected $T$ dependence occurs 
by the impurity effect at very low temperature around $T \sim 2\eta=0.02t$. 
As for the $d$-wave pairing in Fig. \ref{fig:k-T}(b), 
The zero field case shows the $T^2$-behavior~\cite{Schmitt,Hirschfeld1996}, 
which ischaracteristic of 
the line node of the $d_{x^2-y^2}$-wave superconductivity. 
For $H\neq0$, it is modified to 
the $T$-linear dependence~\cite{Suderow,Aubin1999} at low $T$ region. 

It is seen in Fig. \ref{fig:k-T} that 
there exists a crossover temperature $T^\ast$ 
both in the $s$-wave and the $d$-wave pairing cases, 
when we consider the dependence of $\kappa_{xx}(T)$ 
on the magnetic field. 
At lower temperature $T<T^\ast$, $\kappa_{xx}(T)$ increases  
with raising magnetic field.  
However, at higher temperature $T>T^\ast$, $\kappa_{xx}(T)$ decreases 
as a function of $H$.  
It is also noteworthy that $\kappa_{xx}(T)$ shows the 
$T$-linear behavior at $T<T^\ast$, while it deviates from 
$T$-linear at $T > T^\ast$. 
In our parameter, $T^\ast \sim 0.10t $ in the $s$-wave case, and 
$T^\ast \sim 0.06t $ in the $d$-wave case.

\subsection{Position-resolved thermal conductivity}

To understand the abovementioned difference between $T<T^\ast$ and $T>T^\ast$, 
we analyze the position-resolved $\kappa_{xx}({\bf r})$ 
of Eq.(\ref{eq:kappar}), and 
investigate how the local thermal flow contributes to the 
total thermal conductivity. 
The spatial structures of $\kappa_{xx}({\bf r})$ are shown in 
Fig. \ref{fig:k-r-s} for the $s$-wave pairing 
and in Fig. \ref{fig:k-r-d} for the $d$-wave pairing 
from low temperature to high temperature. 
As for the vortex core size, the order parameter recovers at two or 
three lattice sites from the vortex center at low temperatures. 
It is seen for both pairing cases that the heat flows exclusively
at the core region at low $T$. 
It comes from the available low energy excitations around the vortex core.
Around the core region the heat flow extends to the NN vortex direction.  
It is due to the inter-vortex quasi-particle transfer effect.
With increasing $T$, the the contributing region of $\kappa_{xx}({\bf r})$ 
becomes wider around the vortex core and the lines between NN vortices, 
as seen in Figs. \ref{fig:k-r-s}(b) and \ref{fig:k-r-d}(b). 
On the other hand, at higher temperature $T > T^\ast$, 
the dominant contribution comes from the outside region of the vortex core.  
Since $\kappa_{xx}({\bf r})$ is suppressed at the vortex core, 
the vortex core behaves as if it is a scattering center for the 
thermal flow. 
Also along the line connecting NN vortices,  $\kappa_{xx}({\bf r})$ 
is slightly suppressed. 
At further high $T$, the heat flows rather uniformly at all region. 
These higher temperature structure of  $\kappa_{xx}({\bf r})$  comes from 
the contribution of the scattering state at $E > |\Delta|$, 
as discussed later. 

Let us discuss important differences between the $s$-wave and the 
$d$-wave pairing cases. 
At lower temperature $T=0.01t$, 
$\kappa_{xx}({\bf r})$ is well localized
at the core region in the $s$-wave case (Fig. \ref{fig:k-r-s}(a))
compared with the $d$-wave case (Fig. \ref{fig:k-r-d}(a)).
This difference comes from the line node contribution of the 
$d$-wave superconductivity. 
Due to the line node, the low energy quasiparticle states extend outside of 
the vortex core, especially to the $45^\circ $ direction from $a$ and $b$ 
axis of the crystal lattice~\cite{Ichioka96,Ichioka99B}. 
At higher temperature $T=0.09t$,
$\kappa_{xx}({\bf r})$ is well suppressed
at the core region in $d$-wave case (Fig. \ref{fig:k-r-s}(c)),
and it is not too suppressed at the core region in $s$-wave case 
(Fig. \ref{fig:k-r-d}(c)). 
To understand these characteristic behavior of $\kappa_{xx}({\bf r})$, 
we consider the relation 
between the thermal conductivity and the LDOS. 

\begin{figure}[t]
\center{\includegraphics{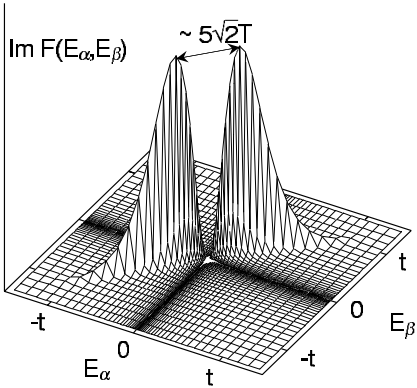}}
\caption{\label{fig:Fab}
$E_\alpha$- and $E_\beta$-dependence of 
the function ${\rm Im }F(E_\alpha,E_\beta)$ of Eq. (\ref{eq:Fab0}). 
In this figure, we set $\eta/t$=0.01 and $T/t$=0.10.
}
\end{figure}

\begin{figure*}[t]
\center{\includegraphics[width=15cm]{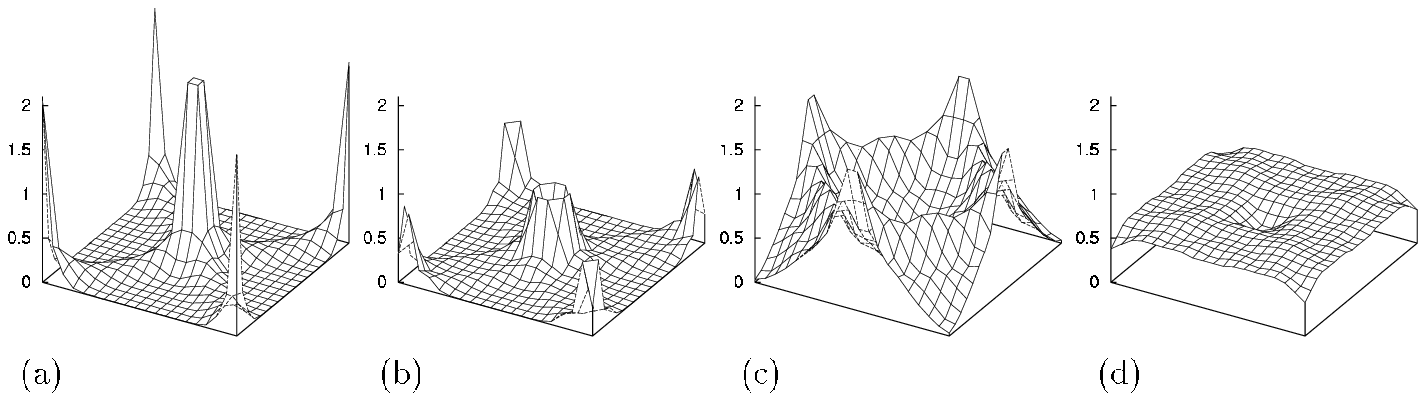}}
\caption{\label{fig:LDOS-E-s}
Local density of states in the $s$-wave pairing case 
at $E/\Delta$=0.34 (a), 0.5 (b), 1.0 (c) and 1.5 (d). 
$H=H_{20}$. 
}
\end{figure*}

\begin{figure*}[t]
\center{\includegraphics[width=15cm]{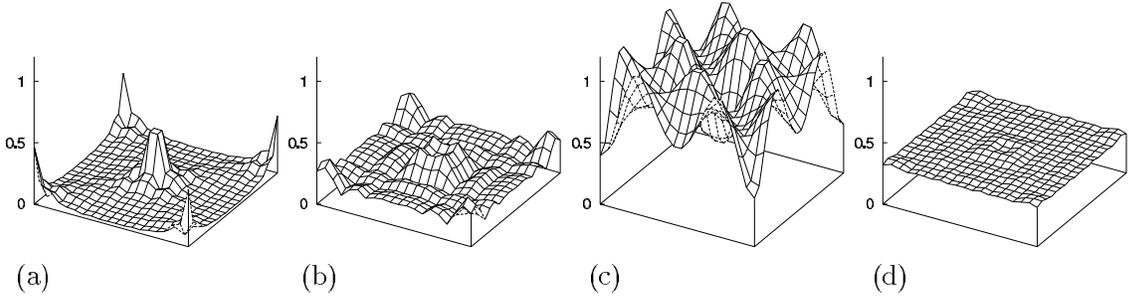}}
\caption{\label{fig:LDOS-E-d}
Local density of states in the $d$-wave pairing case.%
at $E/\Delta$=0 (a), 0.5 (b), 1.0 (c) and 1.5 (d). 
$H=H_{20}$. 
}
\end{figure*}

\subsection{Relation with the local density of states}

In Eq. (\ref{eq:QPP}), temperature dependence comes from 
the function $F({E_\alpha,E_\beta})$ in Eq. (\ref{eq:Fab0}).  
It determines which energy level dominantly contributes to 
the thermal conductivity. 
To see it, we show the $E_\alpha$ and $E_\beta$ dependence of 
${\rm Im }F({E_\alpha,E_\beta})$ in Fig. \ref{fig:Fab}. 
From the figure, we see that ${\rm Im }F({E_\alpha,E_\beta})$ becomes large 
on the line $E_\alpha=E_\beta$. 
Along the line $E_\alpha=E_\beta=E$, we obtain 
\begin{eqnarray} 
{\rm Im }F(E,E) =E^2f'(E).  
\end{eqnarray}
Then, ${\rm Im }F(E,E)$ vanishes at $E=0$, 
and has two peaks at finite energy, which are symmetric with respect to $E=0$. 
The distance of these peaks is about $5\sqrt{2}T$.
It means that the quasiparticle states at these peak energy 
$E \sim \pm 2.5 T$ dominantly contributes to $\kappa_{xx}({\bf r})$. 
At low temperature, the dominant contribution comes from the low energy 
quasiparticle state around the vortex core. 
However, at higher temperature, the scattering states at $E > \Delta(T)$ 
dominantly contribute to the thermal conductivity. 
In this respect, 
thermal conductivity is qualitatively 
different from other physical quantities such as 
electric conductivity, specific heat~\cite{Volovik,Ichioka99A,Wang}, 
nuclear magnetic relaxation time~\cite{TakigawaPRL,TakigawaJPSJ}. 
In these quantities, the quasiparticles at $E\sim 0$ gives largest 
contribution in all temperature regions.
We discuss the electric conductivity at Sec. \ref{El-Cond}.  

At each energy level $E_\alpha$, the contribution to the spatial structure of 
$\kappa_{xx}({\bf r})$ is determined from the spatial distribution of 
the wave functions $u_\alpha({\bf r})$ and $v_\alpha({\bf r})$. 
Then, we show the spatial structure of the LDOS at several energies 
for the $s$-wave pairing in Fig. \ref{fig:LDOS-E-s} and for the 
$d$-wave pairing  in Fig. \ref{fig:LDOS-E-d}. 
In the $s$-wave pairing, the low energy quasiparticle states are bounded 
within the vortex core region, and the quantized energy levels appear 
at half integer energy $E_n = (n + \frac{1}{2})E_\Delta$ 
~\cite{CdGM,HayashiPRL}. 
Here, $E_\Delta$ is the level spacing of the order $\Delta_0^2/E_{\rm F}$.
$\Delta_0$ is the superconducting gap at zero field and 
$E_{\rm F}$ is Fermi energy. 
With the BdG theory at clean limit, since the vortex core radius shrinks to 
the atomic scale with lowering temperature by the Kramer-Pesch 
effect~\cite{Kramer,Pesch}, the quantization effect eminently appears 
at $T \sim 0$.  
Then, there are no states just at $E=0$ because of the small gap by the 
quantization in the $s$-wave pairing. 
At the lowest energy level $E \sim 0.17t$ in 
Fig. \ref{fig:LDOS-E-s}(a), 
the LDOS $N(E,{\bf r})$ has sharp peak at the vortex center. 
It is a bound state in the vortex core. 
At higher energy, the LDOS has peak along a circle around each vortex 
(Fig. \ref{fig:LDOS-E-s}(b)). 
The radius of the circle increases with raising energy. 
We also see the small ridge between NN vortices. 
It is due to the inter-vortex transfer of the low energy bound states. 
At $E \sim \Delta$, the circle of the LDOS peak around the vortex 
overlaps each other (Fig. \ref{fig:LDOS-E-s}(c)). 
At higher energy than $\Delta$, the LDOS are reduced to the 
uniform structure, though the LDOS is slightly suppressed at the 
vortex core region (Fig. \ref{fig:LDOS-E-s}(d)). 
These energy dependence is consistent with the results of 
the quasiclassical calculation~\cite{Ichioka97,IchiokaJS}.  

For the $d$-wave pairing in Fig. \ref{fig:LDOS-E-d}, 
since the low energy states extends outside of the vortex core 
due to the node of the $d_{x^2-y^2}$-wave superconducting gap, 
energy levels becomes 
continuous~\cite{FranzBdG,Franz2000,Wang,TakigawaPRL,TakigawaJPSJ}.
In Fig. \ref{fig:LDOS-E-d}(a), there is a peak of the LDOS at the 
vortex core at $E=0$, which corresponds to the zero-bias peak 
in th spectrum at the vortex center. 
With increasing $E$, the peak of the LDOS is shifted to 
the outside of the vortex, and it is slightly suppressed at 
the vortex center, as shown in Fig. \ref{fig:LDOS-E-d}(b). 
In the $d$-wave case, the large LDOS region shows the fourfold symmetric 
structure instead of the circle structure of the $s$-wave case. 
At $E \sim \Delta$, the LDOS peak around the vortex is shifted 
to the boundary region between vortices (Fig. \ref{fig:LDOS-E-d}(c)). 
At higher energy than $\Delta$, the LDOS are reduced to the 
uniform structure (Fig. \ref{fig:LDOS-E-d}(d)).

\subsection{Energy decomposition of thermal conductivity}

\begin{figure}[t]
\center{\includegraphics[]{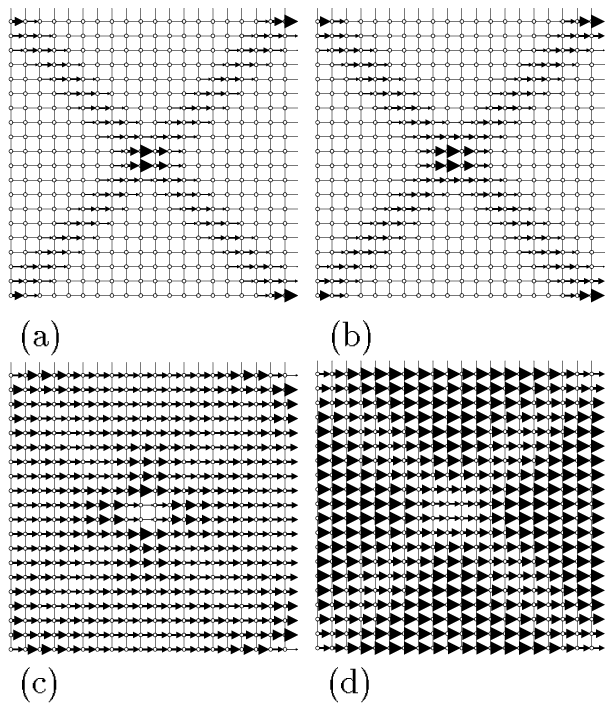}}
\caption{\label{fig:k-r-E-s}
Energy decomposed $\kappa_{xx}({\bf r})$ in the $s$-wave case 
at $T/t=$ 0.02 [(a),(c)] and 0.09 [(b),(d)].
(a) and (b) are for the low energy contribution 
from $|E_\alpha|,|E_\beta|<\Delta(T)/2$. 
(c) and (d) are for the high energy contribution from 
$|E_\alpha|,|E_\beta|>\Delta(T)/2$. 
}
\center{\includegraphics[]{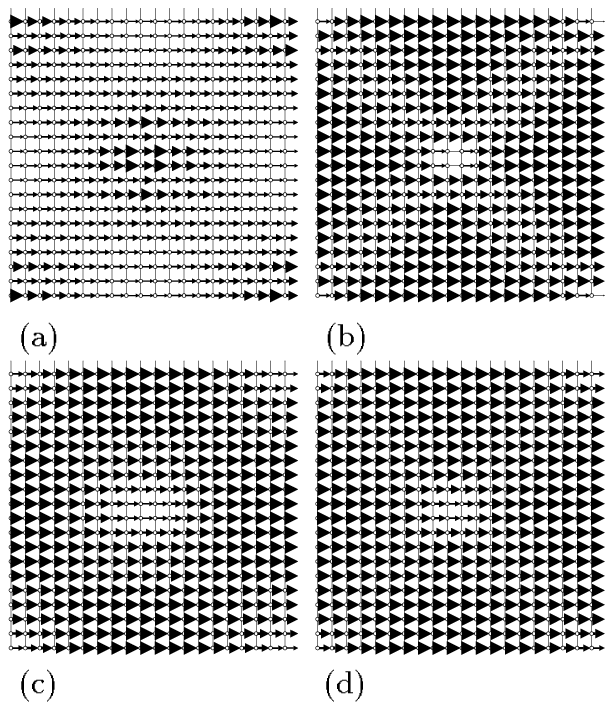}}
\caption{\label{fig:k-r-E-d}
Energy decomposed $\kappa_{xx}({\bf r})$ in the $d$-wave pairing case.
at $T/t=$ 0.02 [(a),(c)] and 0.09 [(b),(d)].
(a) and (b) are for the low energy contribution 
from $|E_\alpha|,|E_\beta|<\Delta(T)/2$. 
(c) and (d) are for the high energy contribution from 
$|E_\alpha|,|E_\beta|>\Delta(T)/2$. 
}
\end{figure}

To discuss the contribution of the LDOS structure 
to the spatial structure of $\kappa_{xx}({\bf r})$, 
We decompose $\kappa_{xx}({\bf r})$ of Eqs. (\ref{eq:QPP}) and 
(\ref{eq:kappar}) into the low energy contribution from 
$|E_\alpha|,|E_\beta|<\Delta(T)/2$ and the high energy contribution from 
$|E_\alpha|,|E_\beta|>\Delta(T)/2$. 
The $s$-wave pairing case is shown in Fig. \ref{fig:k-r-E-s}. 
The upper panels (a) and (b) present the low energy contribution of 
$\kappa_{xx}({\bf r})$, which are localized around the vortex core and 
along the line connecting NN vortices. 
It is because the low energy states for $|E| < \Delta(T)$ are localized 
around the vortex core and there are some inter-vortex quasiparticle transfer 
along the NN vortices direction, as shown in  Fig. \ref{fig:LDOS-E-s}(a) and 
(b). 
The lower panels (c) and (d) show the higher energy contribution of 
$\kappa_{xx}({\bf r})$ from $|E_\alpha|,|E_\beta|>\Delta(T)/2$. 
In this energy range, the wave functions are dominantly located outside 
of the vortex core, as shown in Fig. \ref{fig:LDOS-E-s}(c) and (d).  
Then, the higher energy contribution of $\kappa_{xx}({\bf r})$ is suppressed 
around the vortex core. 
The suppression along the NN vortices directions shown in Fig. 
\ref{fig:k-r-E-s}(d) also reflects the spatial structure of the 
LDOS in  Fig. \ref{fig:LDOS-E-s}(d). 
At low temperature in  Fig. \ref{fig:k-r-E-s}(c), 
spatial structure is determined by the wave function at $E \sim \Delta(T)/2$ 
in Fig. \ref{fig:LDOS-E-s}(b). 
However, at high temperature case in Fig. \ref{fig:k-r-E-s}(d), 
the contribution of the wave function at $E > \Delta(T)$ 
in Fig. \ref{fig:LDOS-E-s}(d) is dominant. 

The $d$-wave pairing case is shown in Fig. \ref{fig:k-r-E-d}.
In the upper panels (a) and (b) for the lower energy contribution, 
$\kappa_{xx}({\bf r})$ is larger around the core and the lines connecting 
NN vortices. 
It is broadly extending around the vortex, compared with the $s$-wave case, 
because the wave functions are also broadly extending around the vortex 
core as shown in Fig. \ref{fig:LDOS-E-d}(a). 
At higher temperature, 
$\kappa_{xx}({\bf r})$ is large outside of the core 
in Fig. \ref{fig:k-r-E-d}(b), 
though LDOS at $E=\Delta(T)/2$ are little localized around the core.
In the lower panels (c) and (d) for the higher energy contribution, 
$\kappa_{xx}({\bf r})$ is large outside of the core at every temperature. 

\begin{figure*}
\center{\includegraphics[width=10cm]{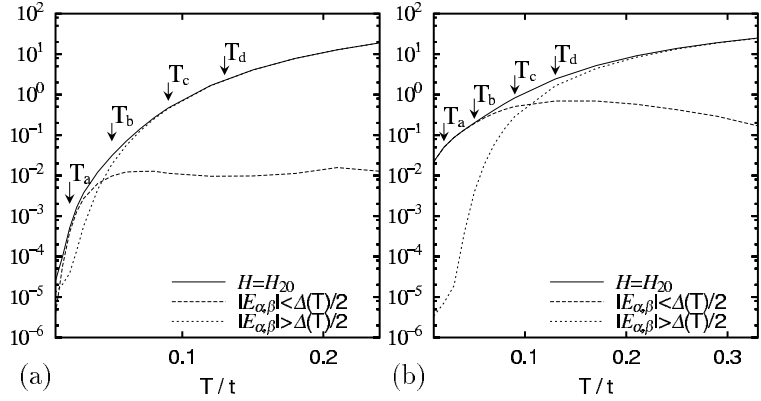}}
\caption{\label{fig:k-w}
Temperature dependence of the energy-decomposed $\kappa_{xx}$. 
The low energy contributions from 
$|E_{\alpha}|, |E_{\beta}| < \Delta(T)/2$ and 
the high energy contributions from 
$|E_{\alpha}|, |E_{\beta}| > \Delta(T)/2$ 
are shown with the total $\kappa_{xx}$(Solid line). 
$H=H_{20}$. 
(a) $s$-wave case. 
(b) $d$-wave case. 
The arrows $T_a$-$T_d$ show $T/t$=0.02, 0.05, 0.09 and 0.13, respectively. 
We show $\kappa_{xx}({\bf r})$ 
in Figures \ref{fig:k-r-s}, \ref{fig:k-r-d}, \ref{fig:k-r-E-s} and \ref{fig:k-r-E-d} at these temperatures.
}
\end{figure*}

Next, we investigate the weight of the energy-decomposed 
contribution for the spatially averaged $\kappa_{xx}$. 
The temperature dependence is presented in Fig. \ref{fig:k-w}.  
For both pairing cases, we can see that the low (high) energy contribution 
is dominant at low (high) temperature. 
The low (high) energy contribution of the $d$-wave pairing case is 
larger (smaller) compared with the $s$-wave pairing case. 
It is because the $d$-wave pairing case has larger DOS at $|E|<\Delta(T)$ 
because the low energy excitation widely extends outside of the vortex core 
region due to the line node of the superconducting gap. 
These DOS difference between the $d$-wave and the $s$-wave pairing cases 
is also shown by the quasiclassical calculation 
(Fig. 18 of Ref. \cite{Ichioka99B}). 
The very low temperature behavior in Fig. \ref{fig:k-w} (a) in the 
$s$-wave pairing at $T < 10^{-2}$ reflects the small gap of the 
quantized energy level in the $s$-wave pairing. 

\section{Electric conductivity}
\label{El-Cond}

\begin{figure}
\center{\includegraphics[]{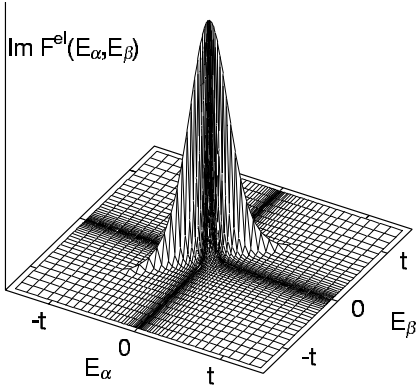}}
\caption{\label{fig:Fab-el}
$E_\alpha$- and $E_\beta$-dependence of 
the function ${\rm Im }F^{el}(E_\alpha,E_\beta)$ of Eq. (\ref{eq:Fab0j}). 
In this figure, we set $\eta/t$=0.01 and $T/t$=0.10.
}
\end{figure}

\begin{figure}
\center{\includegraphics[width=6cm]{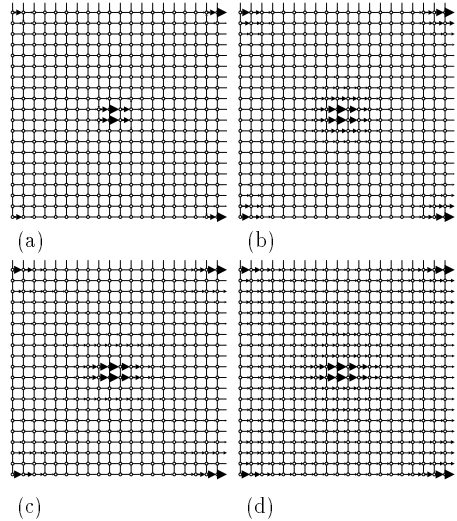}}
\caption{\label{fig:sigma}
The vector plots of the position-resolved electric conductivity 
$\sigma_{xx}({\bf r})$ in the $s$-wave pairing case 
at $T/t$=0.02 (a) and 0.09 (b), and
in the $d$-wave pairing case 
at $T/t$=0.02 (c) and 0.09 (d).  
$H=H_{20}$.  
The vortices are located at the center 
and the four corners in the figure. 
Arrow size is  proportional to flow strength. 
}
\end{figure}

We can also calculate the electric conductivity, if we consider the 
electric current operator instead of the thermal current operator. 
Following the same procedure in Sec. \ref{Linear}, the 
position-resolved electric conductivity is given by 
\begin{eqnarray} && 
\sigma_{xx}({\bf r}_1)
={\rm Im}\{
\frac{\rm d}{{\rm d}\Omega}\frac{1}{N}\sum_{{\bf r}_2} 
Q^{\rm el}_{xx}({\bf r}_1, {\bf r}_2, 
{\rm i}\Omega_n\rightarrow\Omega+{\rm i}0^+ )\}
|_{\Omega\rightarrow0} 
\nonumber \\ && 
\label{eq:sigmar}
\end{eqnarray}
with the correlation function 
\begin{eqnarray}&&%
Q^{\rm el}_{xx}({\bf r}_1 \tau_1,{\bf r}_2 \tau_2) 
=\langle T_{\tau}[j_x({\bf r}_1 \tau_1),j_x({\bf r}_2 \tau_2)]\rangle
\nonumber \\
&&=T\sum_{n}e^{-{\rm i}\Omega_n(\tau_1-\tau_2)}
Q^{\rm el}_{xx}({\bf r}_1,{\bf r}_2,{\rm i}\Omega_n) 
\label{eq:Qxxj}
\end{eqnarray}
of the electric current operator 
\begin{eqnarray} 
{\bf j}({\bf r}_j,\tau) &=& \frac{|e|}{2m}
({\bf P}_j + {\bf P}_{j'}^{\dagger}) 
\nonumber \\&&%
\times
\sum_{\sigma}\psi^{\dagger}_{\sigma}({\bf r}_{j'},\tau)
\psi_{\sigma}({\bf r}_j,\tau)|_{j=j'}.
\end{eqnarray}
Using the wave functions of the BdG equation, we obtain 
\begin{eqnarray} &&
\left. 
\frac{{\rm d}}{{\rm d}\Omega}Q^{\rm el}_{{\bf P}_1{\bf P}_2}
({\bf r}_1,{\bf r}_2,{\rm i}\Omega_n \rightarrow 
\Omega+{\rm i}0^+ ) \right|_{\Omega=0}
\nonumber \\ &&
=-\frac{|e|^2}{4m^2}\sum_{\alpha\beta}F^{\rm el}({E_\alpha,E_\beta})
\nonumber \\ &&
\times
[{{\bf P}_1u_{\alpha}({\bf r}_1)u_{\beta}^\ast ({\bf r}_{1'})
+u_{\alpha}({\bf r}_1){\bf P}_{1'}^{\dagger}u_{\beta}^\ast ({\bf r}_{1'})}
\nonumber \\ &&
-{{\bf P}_1^{\dagger}v_{\alpha}({\bf r}_1)v_{\beta}^\ast ({\bf r}_{1'})
-v_{\alpha}({\bf r}_1){\bf P}_{1'}v_{\beta}^\ast ({\bf r}_{1'})}]
\nonumber \\ &&
\times
[{{\bf P}_2u_{\alpha}({\bf r}_2)u_{\beta}^\ast ({\bf r}_{2'})
+u_{\alpha}({\bf r}_2){\bf P}_{2'}^{\dagger}u_{\beta}^\ast ({\bf r}_{2'})}
\nonumber \\ &&
-{{\bf P}_2^{\dagger}v_{\alpha}({\bf r}_2)v_{\beta}^\ast ({\bf r}_{2'})
-v_{\alpha}({\bf r}_2){\bf P}_{2'}v_{\beta}^\ast ({\bf r}_{2'})}]|_{1=1',2=2'},
\label{eq:Qjj}
\end{eqnarray}
where
\begin{eqnarray} &&
F^{\rm el}({E_\alpha,E_\beta})
\nonumber \\ &&
=\int\frac{d\omega}{2\pi}\int\frac{d\omega'}{2\pi}
\delta_{\eta}(\omega-E_\alpha)\delta_{\eta}(\omega'-E_\beta)
\nonumber \\ &&
\times
[{\rm P}\frac{f(\omega)-f(\omega')}{(\omega-\omega')^2}
+{\rm i}\pi f'(\omega)\delta(\omega-\omega')].
\label{eq:Fab0j}
\end{eqnarray}
We show the $E_\alpha$ and $E_\beta$ dependence of 
${\rm Im }F^{el}({E_\alpha,E_\beta})$ in Fig. \ref{fig:Fab-el}. 
The principal value part of Eq. (\ref{eq:Fab0j}) vanishes by 
the spatial average of $\sigma_{xx}({\bf r}_1)$. 
Then, we consider only the contribution of 
${\rm Im}F^{\rm el}({E_\alpha,E_\beta})$, neglecting 
${\rm Re}F^{\rm el}({E_\alpha,E_\beta})$.  
The function ${\rm Im}F^{\rm el}({E_\alpha,E_\beta})$ becomes large 
for $E_\alpha \sim E_\beta$. 
Along the line  $E_\alpha = E_\beta=E$, 
we obtain ${\rm Im}F^{\rm el}(E,E)=f'(E)$.  
It has maximum at $E=0$ in all temperature region. 
Then, the low energy quasiparticle states around the vortex core 
dominantly contribute to $\sigma_{xx}({\bf r})$ even at higher $T$. 
We present $\sigma_{xx}({\bf r})$ in Fig. \ref{fig:sigma}(a),(b) for 
the $s$-wave pairing case and in Fig. \ref{fig:sigma}(c),(d) for 
the $d$-wave pairing case. 
Even at higher temperature [(b) and (d)], the dominant contribution to 
$\sigma_{xx}$ comes from the vortex core region as in the low temperature 
case [(a) and (c)]. 
This is contrasted with the thermal conductivity case, 
whose dominant contribution 
comes from the outside of the vortex core at high temperatuer, 
as discussed in Sec. \ref{Thermal}. 
Compared to the $s$-wave pairing case[(a) and (b)], 
$\sigma_{xx}({\bf r})$ widely extends 
toward the outside of the vortex core in the $d$-wave pairing case [(c) and (d)]. 
It reflects the extending low energy quasiparticles due to the line node of 
the $d$-wave superconducting gap.

\section{Summary and Discussions}
\label{summary}

We have formulated thermal conductivity in mixed state
based on a microscopic theory of BdG equation and linear
response theory.
The $T$-dependence of thermal conductivity $\kappa_{xx}$ 
for the $s$-wave and the $d$-wave pairing is calculated. 
Their behaviors are analyzed in terms of the position-resolved 
thermal conductivity $\kappa_{xx}({\bf r})$.
And we discuss the relation between $\kappa_{xx}({\bf r})$ and 
the LDOS of the quasiparticles around the vortex. 

There is a crossover temperature $T^\ast$. 
At lower temperature $T<T^\ast$,  $\kappa_{xx}$ is increased with 
raising magnetic field. 
In these temperature region, thermal flow is dominantly carried 
by the low energy quasiparticles around the vortex core and their 
inter-vortex transfer. 
Then, $\kappa_{xx}({\bf r})$ is large around the vortex core and the lines 
connecting NN vortices. 
On the other hand,  at higher temperature $T >T^\ast$, 
$\kappa_{xx}$ is decreased at higher field. 
In these temperature, the contribution to the thermal conductivity 
comes from higher energy quasiparticles including the scattering state 
at $E > \Delta(T)$. 
Therefore, $\kappa_{xx}({\bf r})$ is suppressed at the vortex core region.  
Then, vortex core works as if the scattering center for the thermal flow. 
These contributions from the higher energy states at higher 
temperature is a characteristic of thermal conductivity.  
For other quantities such as electric conductivity, specific heat,
nuclear magnetic relaxation time, the low energy state gives 
largest contribution at all temperature regions. 

The difference between the $s$-wave pairing and the $d$-wave pairing 
comes from the node structure of the superconducting gap. 
The $d$-wave pairing case has larger contribution from the 
low energy quasiparticle states. 
The low temperature distribution of  $\kappa_{xx}({\bf r})$  is broadly 
extending around the vortex in the $d$-wave case, 
because the wave functions are broadly extending around the vortex 
core due to the node structure. 

Our caluclation has reproduced experimental results of the $T$-linear 
behavior at low temperature in the vortex states ($H \neq 0$), and 
the existance of the crossover temperature $T^*$ in the field dependence.
The crossover temperature $T^\ast$ are reported 
both in the conventional $s$-wave superconductor such as Nb 
(Ref. \cite{Sousa}) and in the high-$T_{\rm c}$ superconductors 
such as ${\rm Bi_2 Sr_2 Ca Cu_2 O_8}$ (Ref. \cite{Aubin1999}) and 
in the $f$-wave superconductor ${\rm UPt_3}$ (Ref. \cite{Suderow}). 
These experimental results have been explained as follows. 
At low temperature, vortex assist the thermal flow 
due to the low energy quasiparticle state around the vortex core. 
At higher temperature, vortex behaves as the scattering center 
for the thermal flow.  
Our numerical results of $\kappa_{xx}({\bf r})$ are consistent 
to this picture. 
While the field dependence of $\kappa_{xx}$ is important, 
our calculation cannot examine the contunuous field dependence 
because we consider the field $H=2\phi_0/(aN_r)^2$ depending on 
the size of the unit cell.
At higher temperature, there appears the effect of the $T$-depending 
$\eta$ due to the inelastic scattering by antiferro 
magnetic spin flactuations~\cite{Yu,Hirschfeld1986,Matsukawa}.
For the further extention of our calculation, we will 
examine the effect of the $T$-dependence or the position dependence 
(inside or outside of the vortex core) of the scattering parameter $\eta$.

\section{Acknkowledgement}
The authors thank N.Hayashi for useful discussions.


\end{document}